\documentclass[a4paper,11pt]{article}
\usepackage{pos}
\usepackage{physics}
\usepackage{subcaption}

\title{A Study of Entanglement and Ansatz Expressivity for the Transverse-Field Ising Model using Variational Quantum Eigensolver}
\ShortTitle{Study of Entanglement and Ansatz Expressivity for the TFIM using VQE}

\author*[a]{Ashutosh P. Tripathi}
\author[a]{Nilmani Mathur}
\author[a]{Vikram Tripathi}

\affiliation[a]{Department of Theoretical Physics, Tata Institute of Fundamental Research, \\Homi Bhabha Road, Mumbai 400005, India}

\emailAdd{ashutosh.tripathi@tifr.res.in}
\emailAdd{nilmani@theory.tifr.res.in}
\emailAdd{v.tripathi@theory.tifr.res.in}

\abstract{The Variational Quantum Eigensolver (VQE) is a leading hybrid quantum-classical algorithm for simulating many-body systems in the NISQ era. Its effectiveness, however, depends on the faithful preparation of eigenstates, which becomes challenging in degenerate and strongly entangled regimes. We study this problem using the transverse-field Ising model (TFIM) with periodic boundary conditions in one, two, and three dimensions, considering systems of up to 27 qubits. We employ different ansatzes: the hardware-efficient EfficientSU2 from Qiskit, the physics-inspired Hamiltonian Variational Ansatz (HVA) and HVA with symmetry breaking, and benchmark their performance using energy variance, entanglement entropy, spin correlations, and magnetization.
}

\FullConference{The 42nd International Symposium on Lattice Field Theory (LATTICE2025)\\
2-8 November 2025\\
Tata Institute of Fundamental Research, Mumbai, India\\}

\begin{document}
\maketitle

\section{Introduction}

The Variational Quantum Eigensolver (VQE) is one of the most prominent algorithms in the Noisy Intermediate-Scale Quantum Computing (NISQ) era~\cite{Peruzzo_2014}. It has been applied to study a varieties of quantum many-body systems \cite{Peruzzo_2014,McClean_2016,Tilly_2022, Cao_2022,Stanisic_2022,Zhao_2023}.  
In a VQE method one first considers a suitable wavefunction ansatz for the ground state with a set of variational parameters, and then these parameters are tuned by a classical optimization algorithm to reach the ground state. It can be represented mathematically as,
\begin{equation}
    E_0 = \min_{\vec{\theta}} ~ \langle \psi(\vec{\theta})|H|\psi(\vec{\theta})\rangle,
\end{equation}
where $\vec{\theta}$ is a set of variational parameters. The variational wavefunction ansatz $|{\psi(\vec{\theta})}\rangle$ is represented by a parametric quantum circuit (PQC) in quantum computing. 
Unlike the gate arrangement, since the circuit parameters are tunable, it is assumed that, with a given circuit architecture, the ground state of the system can be achieved with a tuned set of parameter values. Thus, we can say that VQE works with two assumptions: (1) the assumed PQC can represent the ground state, and (2) the classical optimizer can achieve the ground-state parameters.

The Transverse Field Ising Model (TFIM) is a good test model for investigating the efficacy of VQE due to its highly entangled ground state in the small transverse-field regime. 
Although there have been studies of the TFIM using VQE \cite{Wiersema_2020,Jattana_2023,park2024efficientgroundstatepreparation}, here we attempt to benchmark different popular PQCs in terms of expressivity, entanglement, and other observables in order to highlight the role of both the ansatz circuit and the classical optimizer in the VQE framework. The TFIM Hamiltonian is given by,
\begin{equation}
H = J_z \sum_{\langle ij \rangle} \sigma_i^z \sigma_j^z + h_x \sum_i \sigma_i^x,
\end{equation}
where $i$ and $j$ represent positions of the nearest-neighbor spins. The nearest-neighbor interaction strength is represented by $J_z$, and $h_x$ is the transverse magnetic field. We consider $J_z = -1.0$ and periodic boundary conditions throughout this work. We explore this model in one, two, and three dimensions with the above conditions. In particular, we are interested to investigate the entanglement properties with respect to the variation of the strength of the transverse field.

For a finite system, one can qualitatively write the ground state as below:
\begin{itemize}
    \item For $h_x \ll h_c$ ($h_c$ is the critical field for the quantum phase transition):
\begin{equation}
    \begin{split}
        \ket{\psi_0} &= a_0 \left( \ket{00\ldots00} + \ket{11\ldots11} \right) + \cdots, \\
        \ket{\psi_1} &= a_0 \left( \ket{00\ldots00} - \ket{11\ldots11} \right) + \cdots.
    \end{split}
\end{equation}
In this case, the ground state has high entanglement.
    \item For $h_x \gg h_c$:
\begin{equation}
    \ket{\psi_0} = b_0 \ket{++\ldots++} + \cdots.
\end{equation}
In this case, the ground state has low entanglement.
\end{itemize}
Using the variational quantum eigensolver, one can access the ground state and thereby study its entanglement properties. The performance of VQE, however, depends sensitively on the choice of ansatz circuit; an inappropriate ansatz can significantly hinder reliable measurements of quantities such as entanglement entropy. It is therefore essential to assess the effectiveness of different ansatz circuits when applying VQE. In this work, we systematically explore various ansatze and their expressivity in extracting the entanglement properties of the transverse-field Ising model.

\subsection{Exploring various ansatzes and expressivity for VQE}
\begin{figure}[htb]
    \centering
    \begin{subfigure}{0.49\linewidth}
        \includegraphics[width = \textwidth, height= 4.cm]{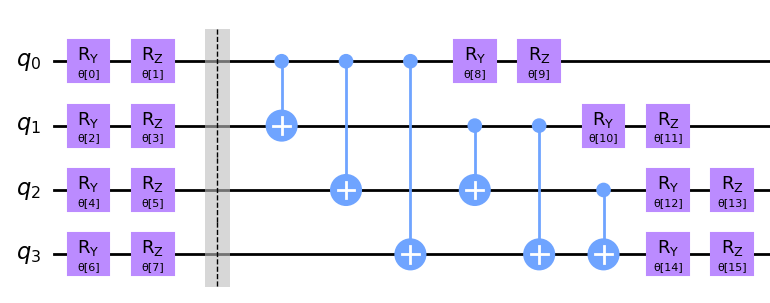}
        \caption{HEA}
    \end{subfigure}
    \begin{subfigure}{0.49\linewidth}
        \includegraphics[width = \textwidth, height= 4.cm]{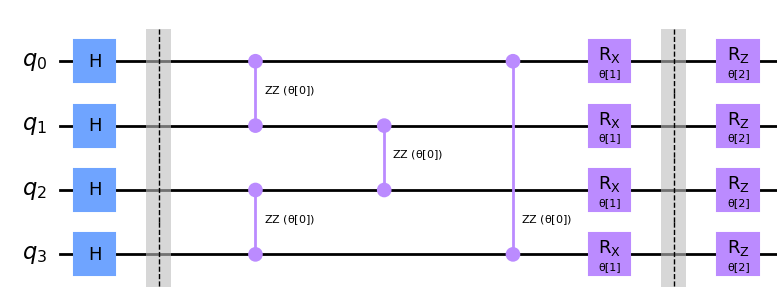}
        \caption{HVA and SB}
    \end{subfigure}
    \caption{HEA and HVA-SB circuit}
    \label{fig:HEA_HVA_circuits}
\end{figure}

We start with three types of ansatz circuits, namely the Hardware-Efficient Ansatz (HEA)~\cite{qiskit2024}, the Hamiltonian Variational Ansatz (HVA)~\cite{Wiersema_2020}, and the Hamiltonian Variational Ansatz with Symmetry Breaking (HVA-SB)~\cite{park2024efficientgroundstatepreparation}. The corresponding parametric quantum circuits (PQCs) are shown in Fig.~\ref{fig:HEA_HVA_circuits}. The circuit in the right panel shows the HVA circuit and the symmetry-breaking layer consisting of $R_z$ gates. The HEA circuit is part of IBM's Qiskit package and is called EfficientSU2. It is built from native hardware gates that are comparatively straightforward to implement. The HVA circuit is constructed by Trotterizing the Hamiltonian terms. The $R_{ZZ}$ gates represent the coupling terms, and the $R_x$ gates arise from the transverse-field term in the TFIM Hamiltonian. In HVA-SB, we add an $R_z$ layer to break the symmetry, providing an overlap state for the Hamiltonian.

\begin{figure}[htb]
    \centering
    \includegraphics[width=0.48\linewidth]{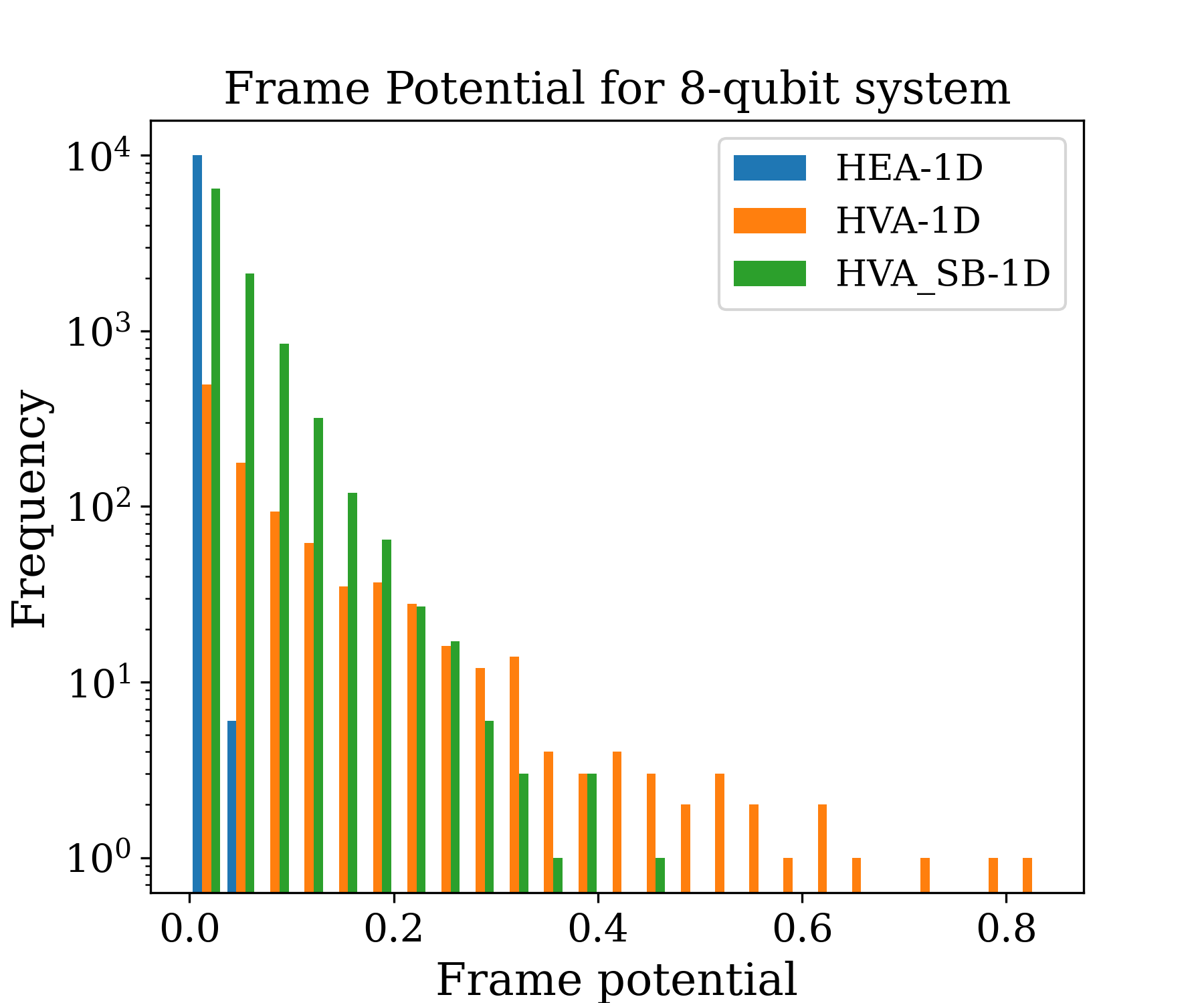}
    \caption{Frame potential calculation}
    \label{fig:frame_potential}
\end{figure}

The expressivity of an ansatz is defined as the ability of a parametric quantum circuit (PQC) to approximate a uniformly random state through variations of its parameters. A common quantitative measure of expressivity is the frame potential associated with a unitary t-design \cite{Sim_2019}, defined as
\begin{equation}
    F_t = \mathbb{E}_{\psi,\phi} \left[ \left| \langle \psi | \phi \rangle \right|^{2t} \right],
\end{equation}
where the expectation value is taken over pairs of states $\ket{\psi}$ and $\ket{\phi}$ generated by randomly sampling the circuit parameters. For highly expressive circuits, the generated states are uniformly distributed over the Hilbert space under uniform sampling of the parameter set, resulting in minimal state overlap i.e. small frame potential value. The histogram shown in Fig.~\ref{fig:frame_potential} represents the distribution of $\left| \langle \psi | \phi \rangle \right|^{2}$ obtained from $10^4$ independent samples. From this distribution, we observe that the HEA yields the lowest frame-potential values, indicating the highest expressivity among the considered ansatzes.

With these ansaztes we now proceed to compute the ground state of TFIM using VQE.

\section{Computational framework}
We perform exact VQE simulations to benchmark ansatz performance using the NVIDIA CUDA-Q simulator \cite{the_cuquantum_development_team_2023_10068206}. Within the VQE framework, we compare the above mentioned three types of variational circuits: the hardware-efficient ansatz (HEA), the Hamiltonian variational ansatz (HVA) tailored to the transverse-field Ising model (TFIM), and a symmetry-breaking (SB) extension of HVA that explicitly breaks the $Z_2$ parity symmetry via additional $R_z$ rotations. Due to its smooth optimization landscape, HEA is optimized using the L-BFGS algorithm, whereas the more rugged landscapes of HVA and SB are handled with the derivative-free COBYLA optimizer.

For each ansatz, we vary the circuit depth (4, 8, 10, and 15 layers) in one- and two-dimensional TFIM, probing the trade-off between expressivity and optimization difficulty. Exact state-vector simulations are performed on NVIDIA GPUs using the CUDA-Q framework, enabling systems of up to 15 qubits in 1D and $4\times4$ lattices in 2D, with exact diagonalization used for benchmarking. From the optimized variational states, we compute ground-state energies and physical observables including magnetization, spin correlations, energy variance, and von Neumann entropy, highlighting the strong dependence of VQE performance on both ansatz structure and optimizer choice. The magnetization,
\begin{equation}
    M = \frac{1}{N}\sum_i \langle \sigma_i^z \rangle,
\end{equation}
vanishes for the exact ground state due to symmetry. Although VQE may yield a finite magnetization in the ordered phase, this effect originates from residual overlap between the ground and first excited states, effectively mimicking spontaneous symmetry breaking in the thermodynamic limit. Since the true ground state exhibits strong correlations, spin-spin correlations provide a more reliable probe of critical behavior. We therefore define, for even $N$, the correlation function
\begin{equation}
    M_{\mathrm{Corr}} = \frac{1}{N/2}\sum_i \langle \sigma_i^z \sigma_{i+N/2}^z \rangle,
\end{equation}
which is well suited for nearly degenerate finite-size systems.

We benchmark the energy accuracy using the energy variance, which is meaningful only when the state is close to the ground state, since the variance vanishes for any exact eigenstate. We define it as
\begin{equation}
Var(E) = \frac{\langle H^2 \rangle - \langle H \rangle^2}{\langle H \rangle^2}.
\label{Eq:var_E}
\end{equation}
It has been shown that the deviation of the variational energy from the exact value scales linearly with $Var(E)$ \cite{Wu_2024}.

We have quantified the entanglement by computing the bipartite entanglement entropy by partitioning the system into subsystems $A$ and $B$. For a pure ground state $\ket{\psi}$ with density matrix $\rho = \ket{\psi}\bra{\psi}$, the reduced density matrix of subsystem $A$ is obtained by tracing out subsystem $B$,
\begin{equation}
    \rho_A = \mathrm{Tr}_B(\rho).
\end{equation}
For a single-qubit subsystem, the reduced density matrix can be written explicitly in terms of Pauli operators, allowing for efficient evaluation. The entanglement entropy is then given by the von Neumann entropy,
\begin{equation}
    S_{EE} = -\sum_i \lambda_i \log \lambda_i,
\end{equation}
where $\{\lambda_i\}$ are the eigenvalues of the reduced density matrix.

Alternatively, the entanglement entropy can be obtained via the Schmidt decomposition, or equivalently by singular value decomposition of the pure state $\ket{\psi}$ in the bipartitioned Hilbert space, providing an equivalent and numerically convenient formulation.

We have further extended our study to three-dimensional TFIM using a Real amplitude ansatz with $R_y$ rotations and evaluate the entanglement entropy per site. While VQE consistently reproduces ground-state energies close to exact results, it systematically underestimates entanglement, reflecting the difficulty of capturing highly correlated states with shallow circuits. This underscores two key limitations: the restricted entangling capability of shallow ansatzes and the sensitivity of convergence to the classical optimizer. Our benchmarking strategy therefore combines highly expressive ansatzes to obtain accurate energy references with HVA circuits constrained to the Hamiltonian subspace, noting that the latter require careful initialization due to their optimizer-unfriendly landscapes.

\section{Results} 

\begin{figure}[htb]
    \centering
    \includegraphics[width=0.48\linewidth]{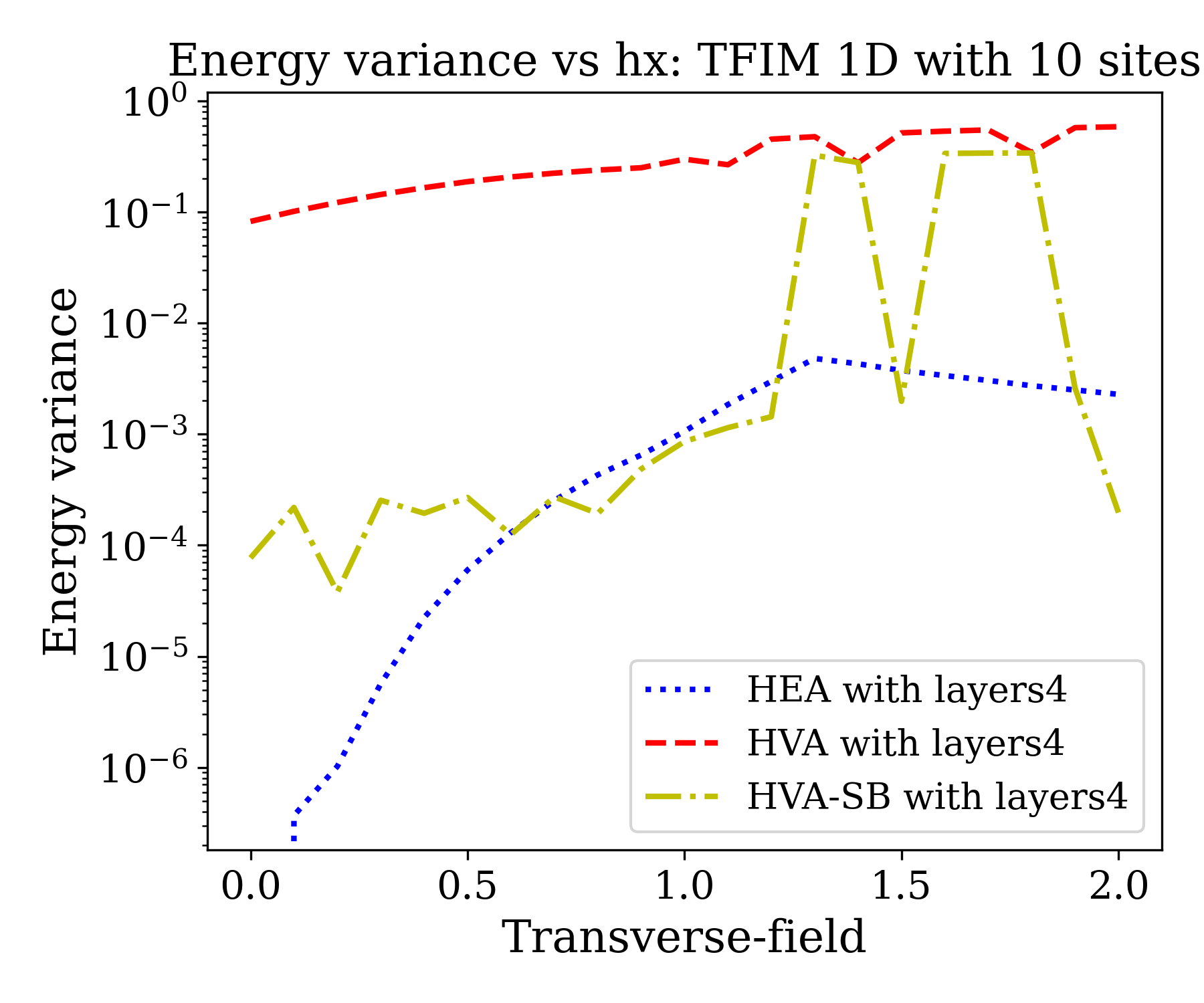}
    \includegraphics[width=0.48\linewidth]{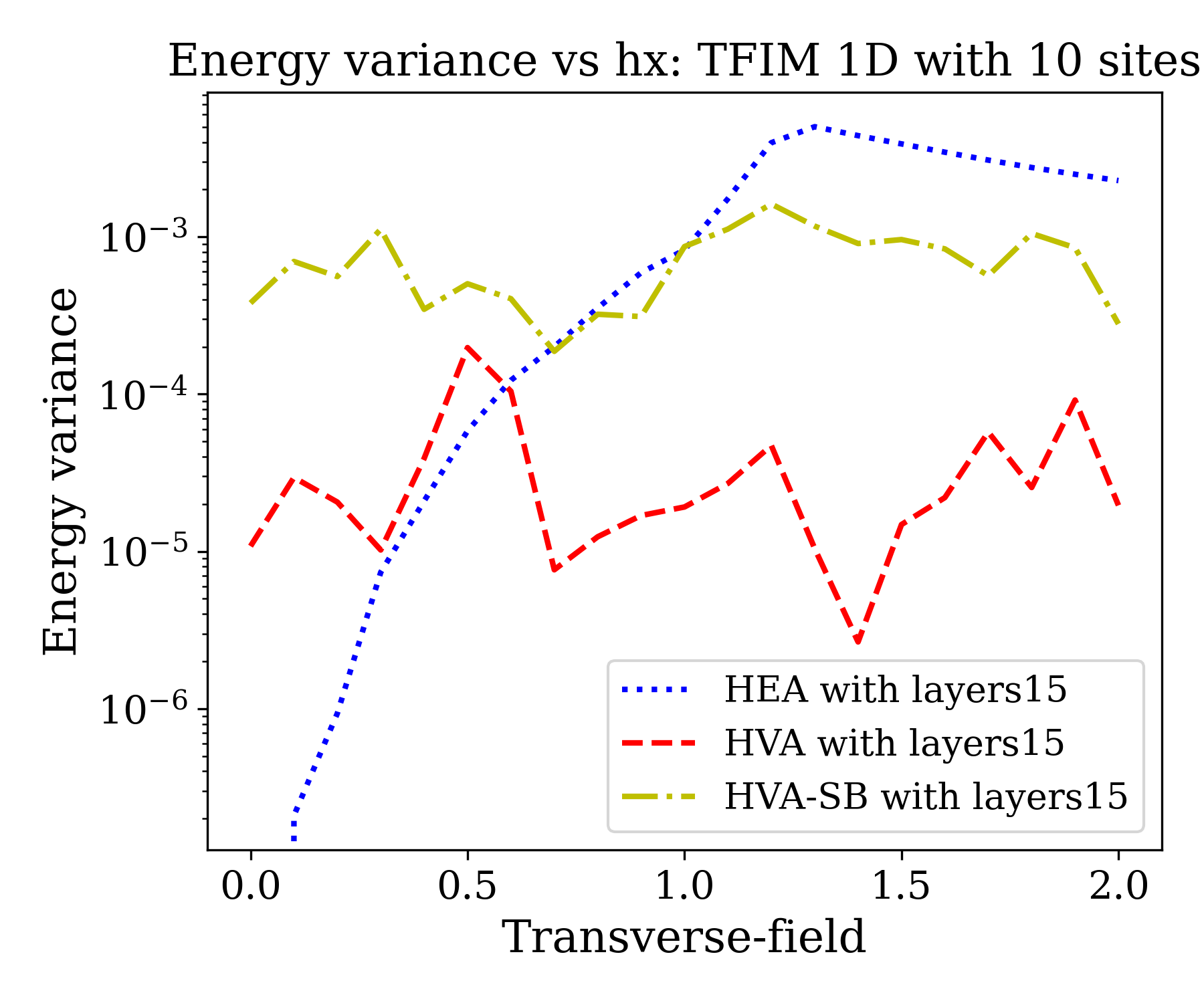}
    \caption{Energy variance vs $h_x$ : TFIM 1-D with 10 sites}
    \label{fig:Var_E_TFIM_1D_10}
\end{figure}

The one-dimensional transverse-field Ising model (TFIM) provides a favorable testbed for VQE studies due to its relative numerical simplicity and the availability of exact solutions. We perform simulations with different system sizes using multiple variational layers. We observe a polynomial scaling of the required number of layers with system size in both one- and two-dimensional lattices. Three classes of ansatz are considered throughout: the hardware-efficient ansatz (HEA), the Hamiltonian variational ansatz (HVA), and the symmetry-breaking extension of HVA (HVA-SB). For the 1D TFIM with ten spins, the energy variance (Eq.\ref{Eq:var_E}) as a function of the transverse field $h_x$ is shown in Fig.~\ref{fig:Var_E_TFIM_1D_10}.

A key distinction among the ansatzes is their parameter count: for a fixed number of layers, HEA contains a number of parameters proportional to the system size, making it significantly more expressive than HVA or HVA-SB. As the circuit depth increases, HEA exhibits a gradual improvement in performance, whereas HVA and HVA-SB display a sharp transition from poor to accurate energy estimation. This behavior arises from the restricted Hilbert space explored by Hamiltonian-based circuits. The inclusion of symmetry-breaking layers in HVA-SB partially relaxes this restriction by allowing access to parity-violating states, leading to smoother improvements compared to pure HVA. The reduced energy variance across different values of $h_x$ for HVA highlights its stability, while HEA shows more irregular behavior.

\begin{figure}[htb]
    \centering
    \begin{subfigure}{0.48\linewidth}
        \includegraphics[width = \textwidth]{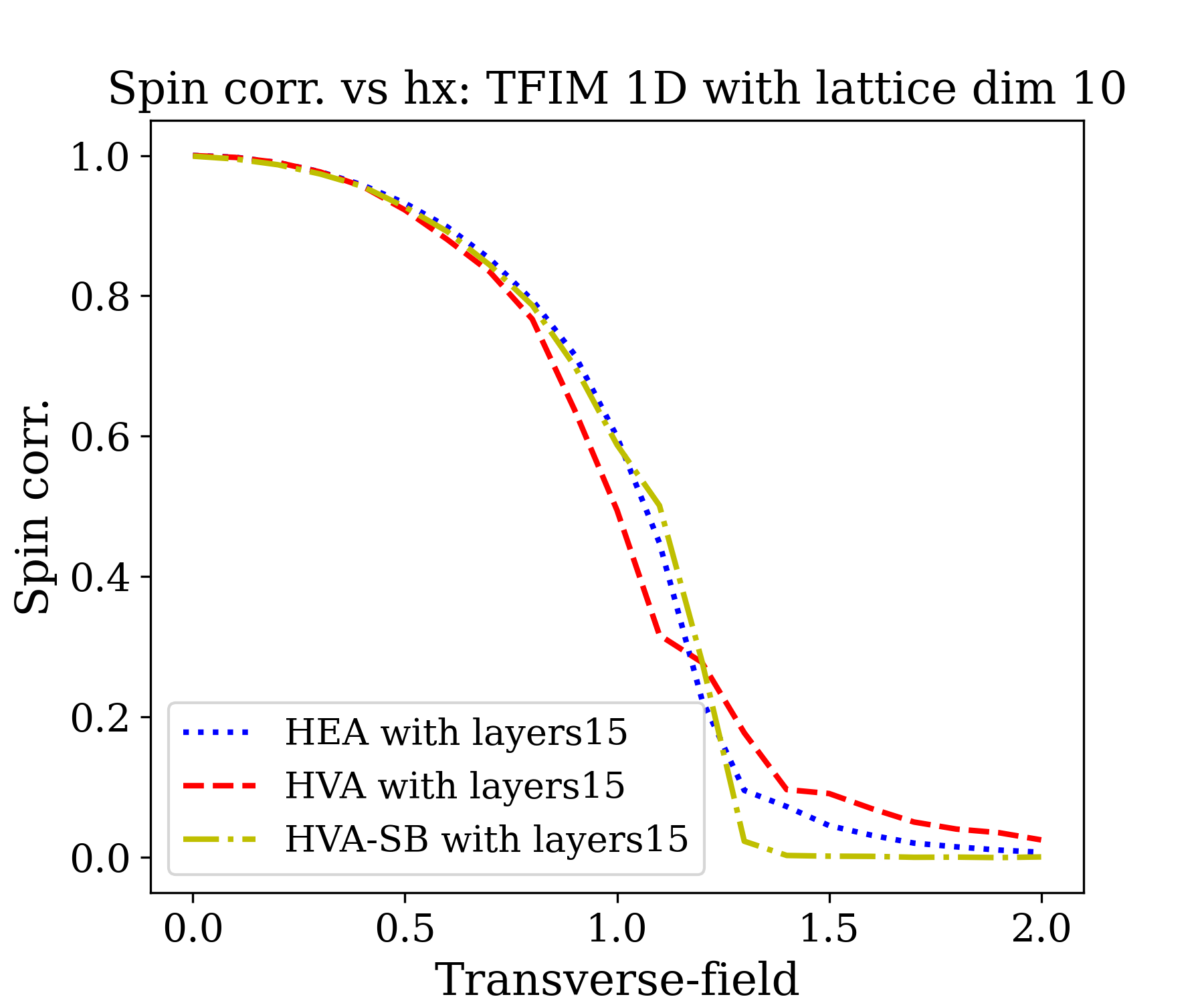}
        \caption{Spin correlation}
    \end{subfigure}
    \begin{subfigure}{0.48\linewidth}
        \includegraphics[width = \textwidth]{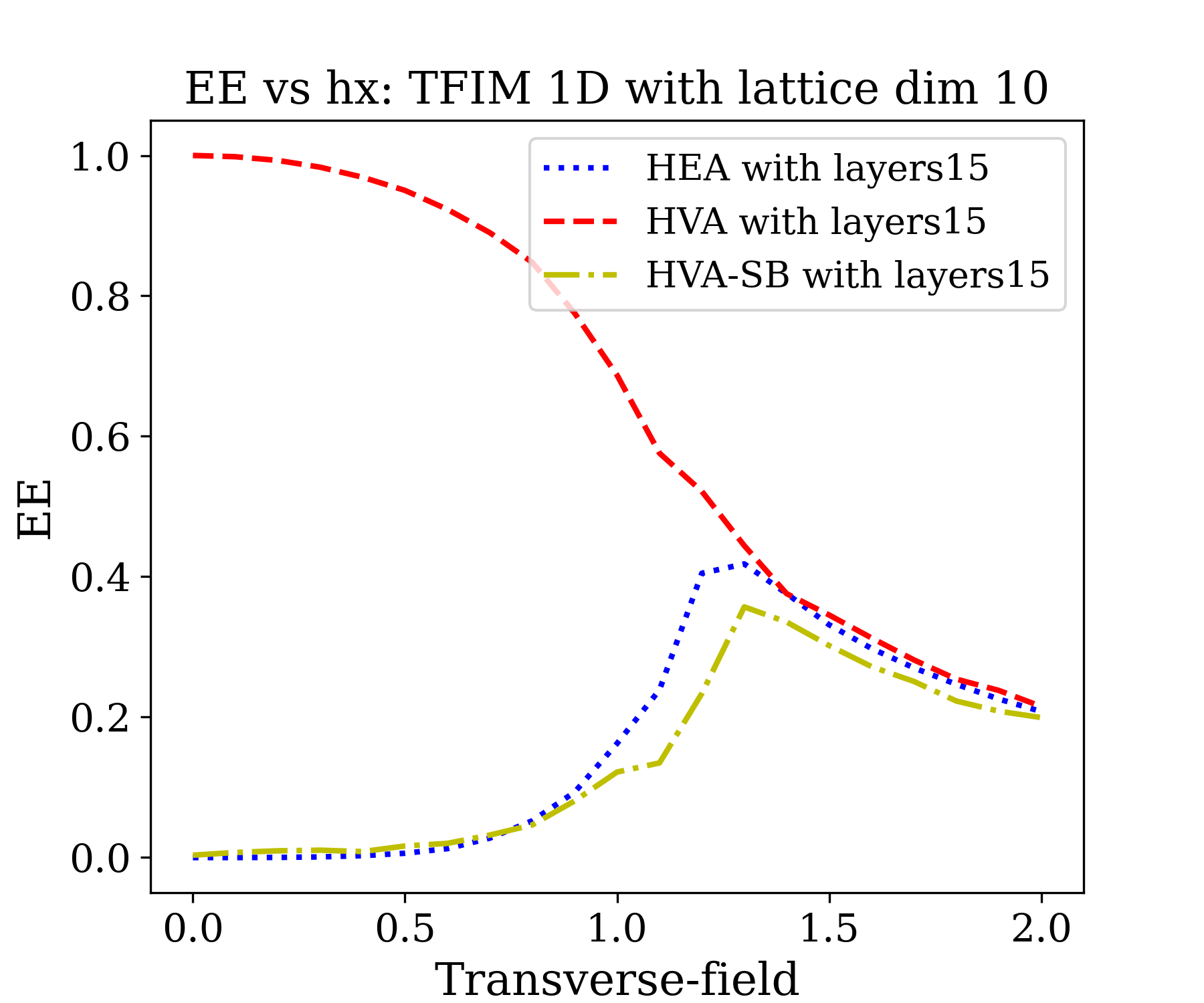}
        \caption{Entanglement entropy}
    \end{subfigure}
    \caption{Spin correlation and Entanglement per site calculation for 10 site 1D system }
    \label{fig:1D_spin_corr_EE}
\end{figure}

Spin correlation and von Neumann entanglement entropy for the 1D system are shown in Fig.~\ref{fig:1D_spin_corr_EE}. The spin correlation exhibits a pronounced change near the critical point, taking approximately half its maximum value. The large single-site EE $=1$ (in units of $\ln 2$) for the HVA is on account of the fact that for finite sizes, the ground state is not spontaneously symmetry-broken. On the other hand, while both HEA nor HVA-SB display a low-field entanglement entropy corresponding to broken symmetry states (which is correct only in the thermodynamic limit), they nevertheless capture qualitative changes in curvature near criticality, underscoring the importance of ansatz selection for resolving phase-transition signatures.

\begin{figure}[htb]
    \centering
    \begin{subfigure}{0.48\linewidth}
        \includegraphics[width = \textwidth]{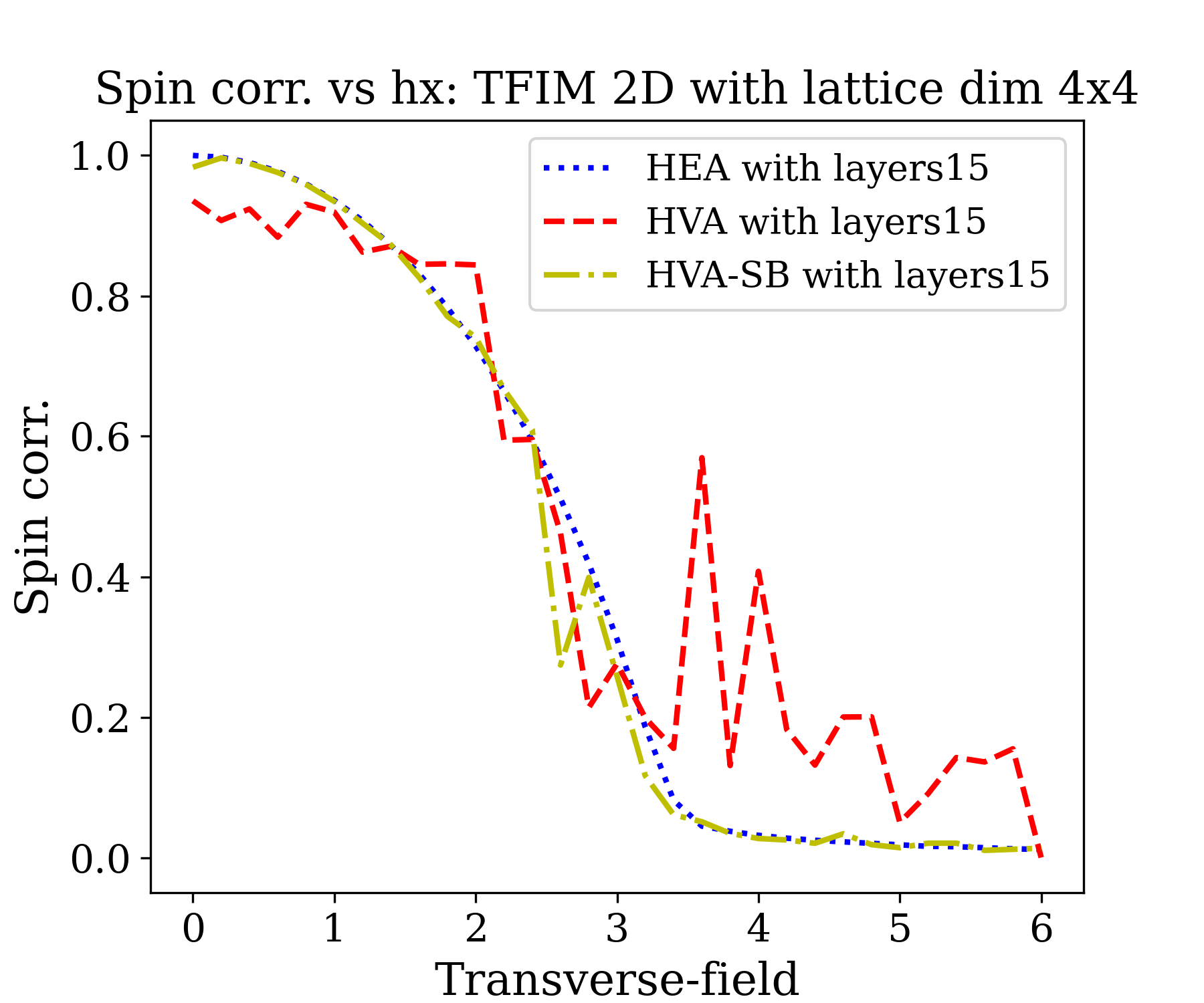}
        \caption{Spin correlation}
    \end{subfigure}
    \begin{subfigure}{0.48\linewidth}
        \includegraphics[width = \textwidth]{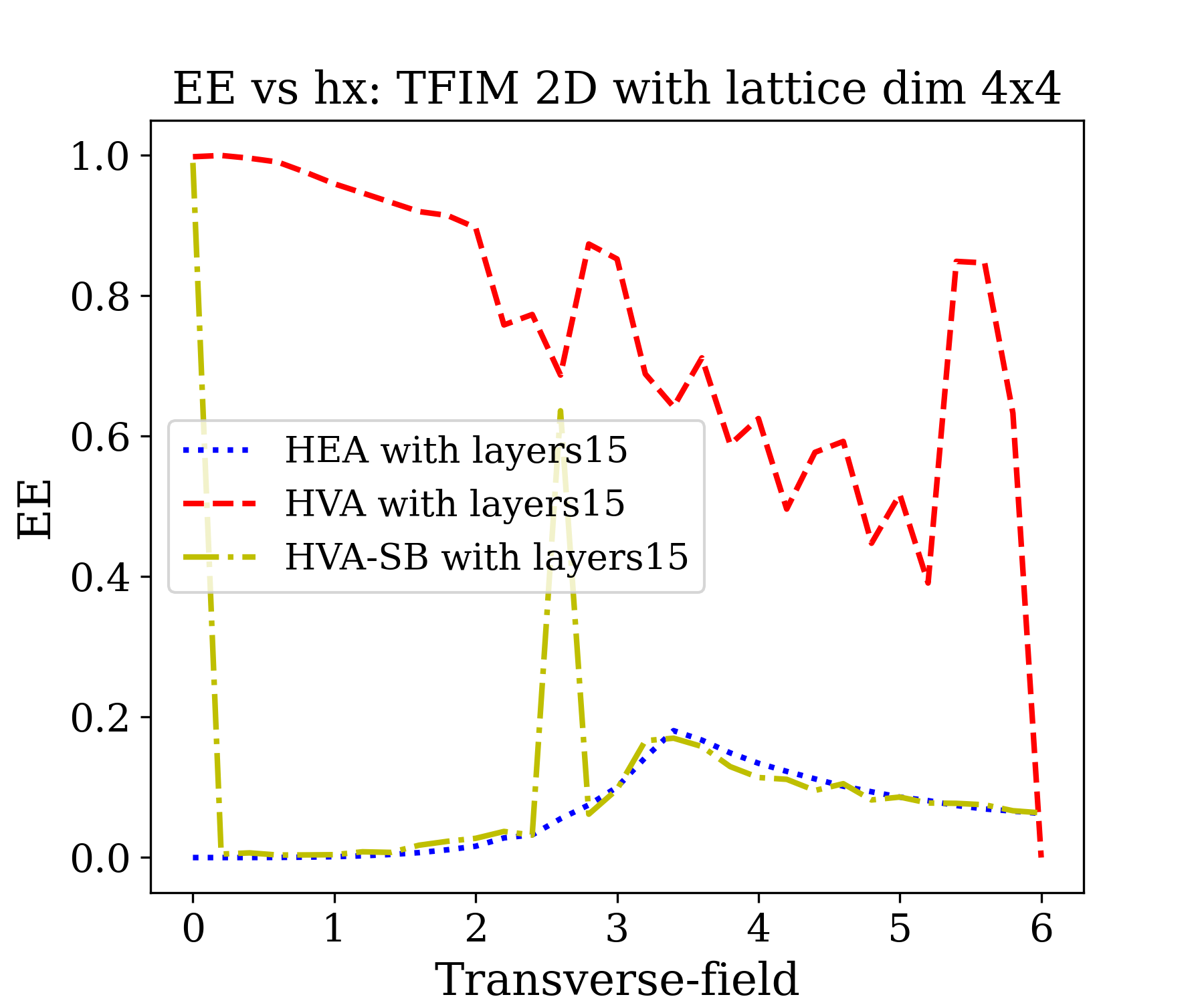}
        \caption{Entanglement entropy}
    \end{subfigure}
    \caption{Spin correlation and Entanglement per site calculation for 2D system of 4x4 dimensions }
    \label{fig:2D_spin_corr_EE}
\end{figure}

Extending the analysis to two dimensions, we apply the same ansatz families to a $4\times4$ TFIM lattice. The optimization becomes noticeably more unstable due to increased connectivity and interactions. As shown in Fig.~\ref{fig:2D_spin_corr_EE}, HVA performs poorly in low-entanglement regimes. HEA underestimates entropy in the highly entangled low-field regions on account of the tendency to pick spontaneously symmetry broken states as we observed above for the 1D case. We expect the symmetry-broken low-field states to correctly describe the ground states in the thermodynamic limit. Different initial parameter choices lead to distinct local minima, complicating quantitative comparisons. Both spin correlation and entanglement entropy clearly signal a phase transition near the critical field $h_x \approx 3.0$.

\begin{figure}[htb]
    \centering
    \begin{subfigure}{0.48\linewidth}
        \includegraphics[width=\linewidth]{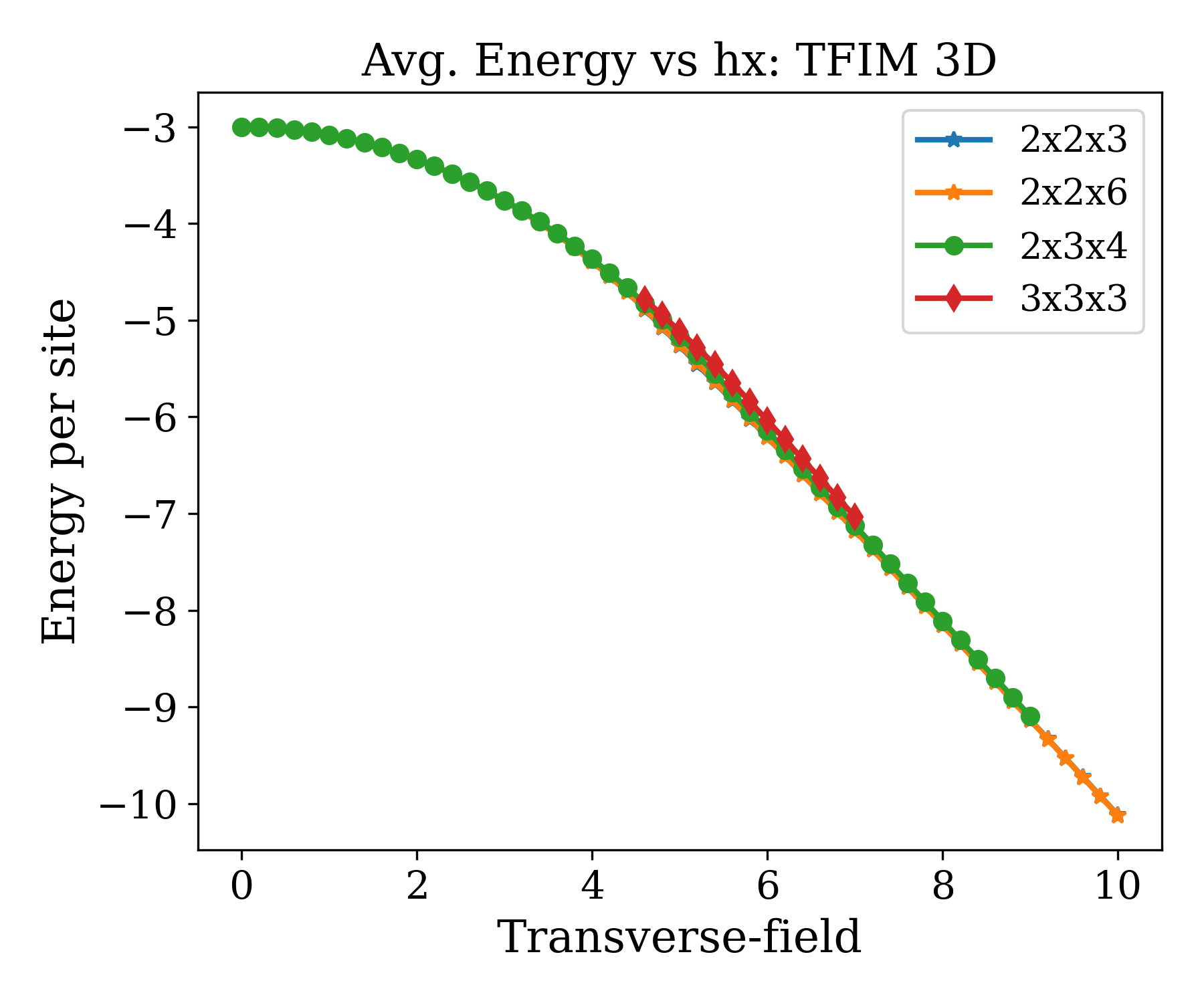}
        \caption{}
        \label{fig:3_d_Energy}
    \end{subfigure}
    \begin{subfigure}{0.48\linewidth}
        \includegraphics[width=\linewidth]{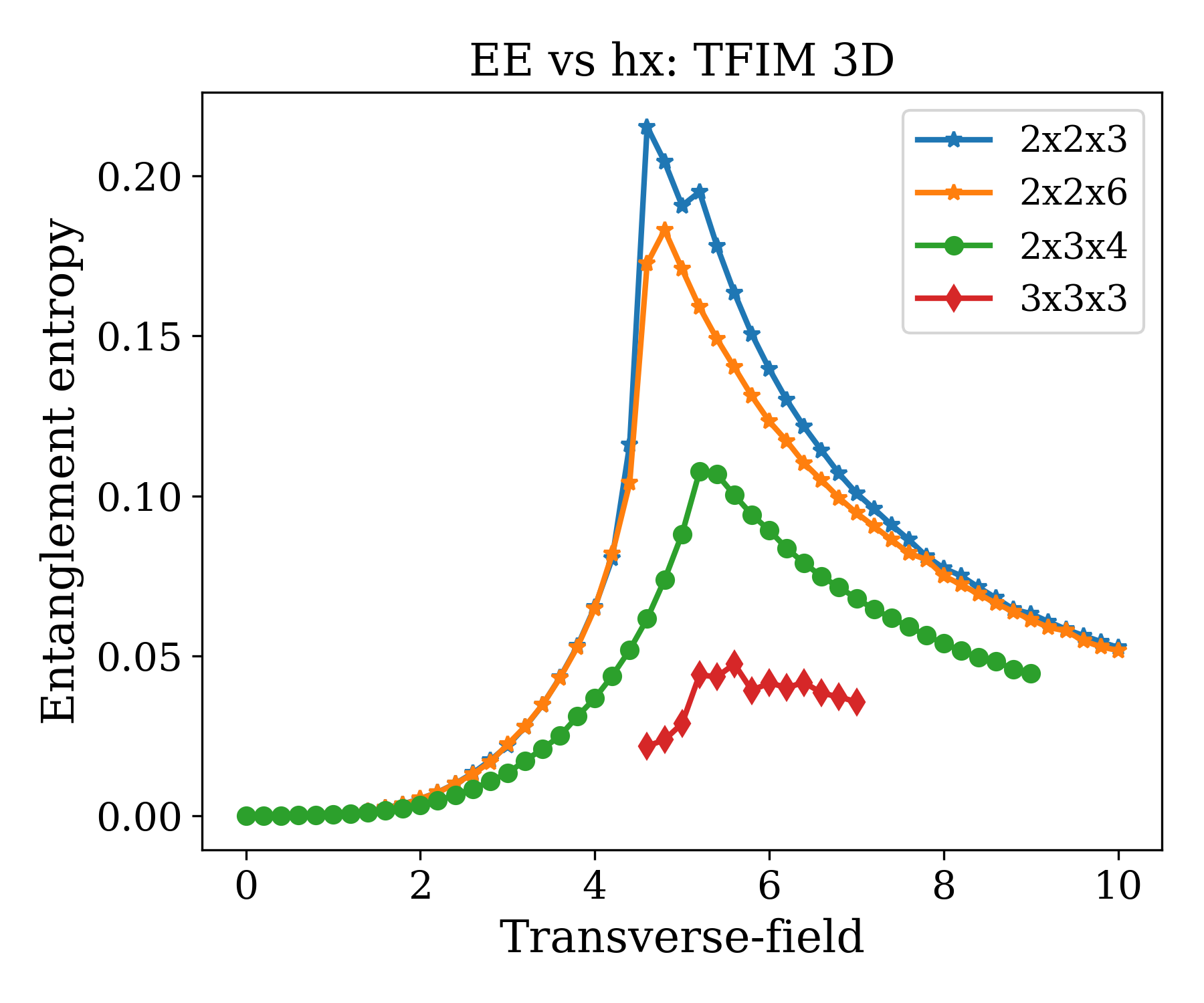}
        \caption{}
        \label{fig:3_d_EE}
    \end{subfigure}
    \caption{ Average energy and entanglement entropy of the 3-d systems for different lattice sizes.}
\end{figure}

In three dimensions, optimization challenges further intensify. Despite this, ground-state energies can still be obtained with reasonable accuracy using a modified HEA. By removing $R_z$ rotations, the ansatz is restricted to real amplitudes, reducing the number of parameters while preserving a smooth optimization landscape. Additionally, initializing parameters using optimized values from nearby transverse-field points significantly reduces optimization time. As shown in Figs.~\ref{fig:3_d_Energy} and \ref{fig:3_d_EE}, the ground-state energy per site remains approximately independent of lattice size, while the entanglement entropy continues to exhibit clear signatures of critical behavior.

\section{Conclusion}

We have performed a comparative study of the HEA, HVA, and HVA-SB ansatzes using the example of transverse-field Ising model, highlighting a clear trade-off between circuit expressivity and optimization performance. We observe that increasing the number of layers leads to only a gradual improvement in HEA, whereas both HVA and HVA-SB show a sudden transition from poor to accurate ground-state energy estimation. This behavior reflects the advantage of restricting the variational circuit to a Hamiltonian-inspired subspace. The role of symmetry breaking in HVA-SB is particularly important in accessing low-entanglement overlap states (overlap of the nearly degenerate ground state), which improves the approximate ground-state energy while illustrating how the ansatz structure influences the ability to reach the true ground state. Signatures of critical behavior are qualitatively visible in the decay of spin correlations near the transition point and in the corresponding behavior of the entanglement entropy. In higher-dimensional simulations, the increased interaction connectivity and stronger entanglement make the optimization more challenging. For finite size lattices, HEA underestimates entanglement in the low-field phase (and tends to select symmetry broken states that correctly capture ground state properties only in the thermodynamic limit), while HVA performs poorly in the low-entanglement region.

Overall, our results show that expressivity and optimization stability are competing requirements in VQE. In future work, adaptive ansatz strategies and machine-learning-based optimizers can be explored to further mitigate these limitations and improve scalability to larger and more strongly correlated systems.

\section{Acknowledgment}
 
We acknowledge the discussion with Debasish Banerjee on the topic of entanglement entropy and variational simulation.
This work is supported by the Department of Atomic Energy, Government of India, under
Project Identification Number RTI 4002. Simulation were carried out on the Guppy and Grouse
clusters of the Department of Theoretical Physics, TIFR. We used Qskit\cite{qiskit2024} and NVIDIA CUDA-Q\cite{the_cuquantum_development_team_2023_10068206} python packages.
\bibliographystyle{JHEP}
\bibliography{VQE_TFIM_POS}

\end{document}